\documentclass{ifacconf}

\usepackage{graphicx}      
\usepackage{natbib}        

\usepackage{amssymb}

\renewcommand{\i}{{\mathrm{i}}}
\newcommand\R{{\mathbb{R}}}

\newcommand{\x}{\mathbf{x}}

\begin{document}
\begin{frontmatter}

\title{Small frequency approximation of (causal) dissipative pressure waves}

\author[First]{R. Kowar} 

\address[First]{Department of Mathematics, University of Innsbruck,
   Technikerstrasse 21a/2, A-6020,Innsbruck, Austria \\ (e-mail: richard.kowar@uibk.ac.at)}

\begin{abstract}                

In this paper we discuss the problem of small frequency approximation of the causal dissipative 
pressure wave model proposed in~\cite{KoScBo:11}. We show that for appropriate situations 
the Green function $G^c$ of the causal wave model can be approximated by a noncausal Green 
function $G_M^{pl}$ that has frequencies only in the small frequency range $[-M,M]$ 
($M\leq 1/\tau_0$, $\tau_0$ relaxation time) and obeys a power law. 
For such cases, the noncausal wave $G^{pl}_M$ contains partial waves propagating arbitrarily 
fast but the sum of the noncausal waves is small in the $L^2-$sense. 
\end{abstract}


\end{frontmatter}

\section{Introduction: frequency framework}

In the first section we summarize the general framework of frequency driven wave dissipation. 

From the mathematical point of view, in case of a \emph{homogeneous} and \emph{isotropic} 
medium, dissipative pressure waves can be modeled by (cf. e.g.~\cite{KoScBo:11})
$$
     p(\x,t) = (G *_{\x,t} f)(\x,t)  \qquad \x\in\mathbb{R}^3,\,t\in\R
$$
with 
\begin{equation}\label{initcond0}
       p(\cdot,t)=0  \quad \mbox{and}\quad   \frac{\partial p}{\partial t}(\cdot,t)=0 
\quad \mbox{for}\quad t<0\,,
\end{equation}
where $G$ denotes a \emph{distribution} (\emph{Green function}), $*_{\x,t}$ denotes the space-time 
convolution and $f$ denotes a \emph{forcing term} (\emph{source term}) which models wave generation. 
We note that $f$ is the same forcing term as in the absence of dissipation as long as~(\ref{initcond0}) 
holds. (This is not true if $p(\x,t)$ is calculated for $t\geq t_0$ ($t_0>0$) from initial data at 
$t_0$.)

\subsection{The Green function}

The Green function can be modeled by
\begin{equation}\label{green}
    \hat G(\x,\omega) 
       = \frac{e^{-\beta_*(|\x|,\omega)}}{4\,\pi\,|\x|}\,e^{\i\,\frac{\omega}{c_0}\,|\x|}
   \qquad\quad \x\in\R^3,\,\omega\in\R\,, 
\end{equation}
where $\Re(\beta_*(|\x|,\omega)) > 0$ is even in $\omega$ and $\Im(\beta_*(|\x|,\omega))$ is 
odd in $\omega$. Here $\hat G(\x,\omega)$ denotes the Fourier transform of $ G(\x,t)$ with 
respect to time $t$. The last two conditions guarantee that $G(\x,t)$ is real-valued. 
We focus on the cases 
\begin{equation}\label{beta}
      \beta_*(\x,\omega) = \alpha_*(\omega)\,|\x| \,.
\end{equation}
We call $\alpha_*=\alpha_*(\omega)$ and $\alpha=\Re(\alpha_*(\omega))$ the 
\emph{attenuation-dispersion law} and the \emph{attenuation law}, respectively.

\subsection{The wave equation}

The above dissipative wave model satisfies the following 
\emph{integro-differential equation} 
\begin{equation}\label{waveeq}
   \nabla^2 p(\x,t) 
         - \left(D_* + \frac{1}{c_0}\,\frac{\partial p}{\partial t} \right)^2 p(\x,t) 
         = -f(\x,t)\,,
\end{equation}
where $D_*$ denotes the time convolution operator 
$$
  \mathcal{F}\{D_*(g)\}(\omega) 
       = \frac{1}{\sqrt{2\,\pi}}\,\alpha_*(\omega)\,\mathcal{F}\{g\}(\omega)
$$
defined for appropriate functions $g=g(t)$. Here $\mathcal{F}\{g\}$ denotes the Fourier transform 
of $g$ with respect to time.  

More details about attenuation-dispersion laws and respective wave equations can be found in 
e.g.~\cite{NaSmWa:90},~\cite{Sz:95},~\cite{WaHuBrMi:00},~\cite{ChHo:04},
~\cite{PaGr:06},~\cite{KeMcMe:08} and~\cite{KoSc:12}.

\section{Attenuation-dispersion laws and causality}

\subsection{What do we mean by causality?}

Let $c_F<\infty$ denote the (constant) speed of the wave front of the wave $G$.  
Causality requires that \emph{the wave front initiated at the origin 
$\mathbf{0}$ at time $t=0$ arrives at position $\x\not=\mathbf{0}$ not before the time period 
$T(|\x|)=\frac{|\x|}{c_F}$ is over.} 
Mathematically this is equivalent to the following \emph{causality condition}: 
for each $c_1\geq c_F$ we have 
\begin{equation}\label{causrequ}
      G\left(\x,t+\frac{|\x|}{c_1}\right) = 0\qquad \mbox{if}\qquad 
             t<0 \;\; \mbox{and} \;\; |\x|\not=0\,.
\end{equation}
Various dissipative wave models are analysed with respect to causality 
in~\cite{KoSc:12}. In particular, it is shown that~(\ref{beta}) 
implies $c_F=const.$ if $c_F <\infty$. See also~\cite{KeMcMe:08}.

\subsection{Attenuation-dispersion laws ($\gamma\in (1,2]$)}

In this paper we consider the \emph{attenuation-dispersion laws} 
\begin{equation}\label{alpha*} 
   \alpha_*^c(\omega) 
      = \frac{\alpha_1(-\i\,\omega)}{c_0\,\sqrt{1+(-\i\,\tau_0\,\omega)^{\gamma-1}}} 
\end{equation}
and 
\begin{equation}\label{alpha*pl} 
       \alpha_*^{pl}(\omega) 
                = a_1\,\frac{(-\i\,\omega)^\gamma}{\cos(\gamma\,\pi/2)} 
                        + a_2\,(-\i\,\omega)
\end{equation}
for $\omega \in\mathbb{R}$ with $\gamma\in (1,2]$, $c_0\in (0,\infty)$ and  
$\alpha_1,\,\tau_0,\,a_1>0$. 
$\alpha^{pl}=\alpha_*^{pl}(\omega)$ is called the \emph{frequency power law}. 

Let
\begin{equation}\label{a1a2}
     a_1 = \frac{\alpha_1\,\tau_0^{\gamma-1}}{2\,c_0}\,|\cos(\gamma\,\pi/2)|
\quad\mbox{and}\quad
     a_2 = \frac{\alpha_1}{c_0} \,,
\end{equation}
then it follows from 
$$
      (-\i\,\omega)^\gamma = |\omega|^\gamma\,
        \left\{\cos\left(\gamma\,\frac{\pi}{2}\right) 
          - \i\,\,\sin\left(\gamma\,\frac{\pi}{2}\right)\mbox{sgn}(\omega)\right\}
$$
that 
$$
\begin{array}{l}      
   \alpha^c(\omega)=\Re( \alpha_*^c(\omega)) \approx a_1\,|\omega|^\gamma = \alpha^{pl}(\omega)\,,\\ 
   \Im( \alpha_*^c(\omega)) \approx -a_1\,\tan\left(\gamma\,\frac{\pi}{2}\right)
          \,\mbox{sgn}(\omega)\,|\omega|^\gamma 
          - a_2\,\omega
\end{array}
$$
for $|\tau_0\,\omega|^{\gamma-1} << 1$. 
We see that $\alpha_*^{pl}(\omega)$ is a good approximation of $\alpha_*^c(\omega)$ for 
\emph{sufficiently small} frequencies. 
For the case $\tau_0=10^{-6}\,\mu\,s$ (liquid), the small frequency 
range condition $|\tau_0\,\omega|^{\gamma-1}\leq 0.1$ is visualized in Table~\ref{tb:sfcond}.

\begin{table}[hb]
\begin{center}
\caption{Visualization of small frequency range condition $|\tau_0\,\omega|^{\gamma-1}\leq 0.1$ 
for $\tau_0=10^{-6}\,\mu\,s$.}
\label{tb:sfcond}
\begin{tabular}{ccc}
$\gamma$ & range                       & bound $M$\\ \hline
$1.1$    & $|\omega|\leq 10^{-4}\,MHz$  & $10^{-4}\,MHz$ \\
$1.5$    & $|\omega|\leq 10^{+4}\,MHz$  & $10^{+4}\,MHz$ \\ 
$2.0$    & $|\omega|\leq 10^{+5}\,MHz$  & $10^{+5}\,MHz$ \\ \hline
\end{tabular}
\end{center}
\end{table}

\subsection{Two results on causality}

In~\cite{KoSc:12}, it is proven that the speed $c_F$ of the wave front of $G(\x,t)$ with  
$\alpha_*^c$ defined as in~(\ref{alpha*}) satisfies $c_F\leq c_0$ and in~\cite{Ko:10} 
it is proven that $c_F= c_0$. 
Moreover, it is shown that $G(\x,t)$  with $\alpha_*^{pl}$ defined as 
in~(\ref{alpha*pl}) with $a_2=0$ has not a bounded wave front speed. The same 
result is true for $a_2\not=0$.

Loosely speaking, causality restricts the growth of $\alpha(\omega)$ and consequently 
$\omega\mapsto\hat G(\x,\omega)$ must not decrease too fast for fixed 
$x\in\R^3\backslash\{\mathbf{0}\}$. If $\omega\mapsto\hat G(\x,\omega)$ decreases too fast, 
then it behaves like the truncated Green function 
\begin{equation}\label{defhatGM}
   \hat G_M(\x,\omega) := \hat G(\x,\omega)\,\chi_{[-M,M]}(\omega) , 
\end{equation}
for which causality condition~(\ref{causrequ}) cannot hold. Here 
\begin{equation}\label{defchi}
  \chi_{[-M,M]}(\omega)
  := \left\{
         \begin{array}{ll}
          1 \,& \mbox{if $\omega\in [-M,M]$} \\
          0                 & \mbox{otherwise}
          \end{array}
       \right. \,
\end{equation}
denotes the \emph{characteristic function} of $\omega$ on $[-M,M]$.

\section{Small frequency approximation}

We now derive two theorems that permit to estimate the quality of  
small frequency approximations. Let 
\begin{itemize}
\item $G^c(\x,t)$ denote the Green function with $\alpha_*^c$ defined as 
      in~(\ref{alpha*}),

\item $G^{pl}(\x,t)$ denote the Green function with $\alpha_*^{pl}$ defined 
      as in~(\ref{alpha*pl}),

\item $\hat G_M(\x,\omega)$ be defined as in ~(\ref{defhatGM}) and 

\item $G_M(\x,t) := \mathcal{F}^{-1}\{\hat G_M\}(\x,t)$. 
\end{itemize}

\begin{thm}\label{theo:theo1}  
Let $M>0$ and $A_0,\,A_1,\,A_3>0$ be such that
\begin{equation}\label{relb1b2}
  \alpha(M) + A_0\,|\omega-M|  
     \leq  \alpha^c(\omega) 
     \leq A_1\,\omega^2 + A_2\,|\omega|   
\end{equation}
holds for $|\omega|>M$. For $|\x|\not=0$ it follows that 
$$ 
   \frac{\|G^c(\x,\cdot) - G^c_M(\x,\cdot)\|_{L^2}}{\|G^c(\x,\cdot) \|_{L^2}} 
   \leq \sqrt[4]{\frac{2\,A_1}{\pi\,A_0^2}}\,
            \frac{e^{-\left(\alpha(M)+\frac{A_2^2}{4\,A_1}\right)\,|\x|}}{\sqrt[4]{|\x|}} \,.
$$
\end{thm}

\begin{pf}
$\omega\mapsto\hat G^c(\x,\omega)$ and $\omega\mapsto\hat G^c_M(\x,\omega)$ 
are square integrable and thus their inverse Fourier transforms $t\mapsto G^c(\x,t)$ and 
$t\mapsto G^c_M(\x,t)$ are square integrable, too. Because of the 
\emph{Plancherel-Parseval equality}, it follows  
\begin{equation}\label{resofPPE}
      \frac{\|G^c(\x,\cdot) - G^c_M(\x,\cdot)\|_{L^2}}{\|G^c(\x,\cdot) \|_{L^2}} 
   = \frac{\|\hat G^c(\x,\cdot) - \hat G^c_M(\x,\cdot)\|_{L^2}}{\|\hat G^c(\x,\cdot) \|_{L^2}} \,. 
\end{equation}
From~(\ref{relb1b2}) and 
$$
   \int_{-\infty}^\infty\exp\{-\omega^2/2\}\,\d \omega =\sqrt{2\,\pi}\,,
$$ 
it follows 
$$
\begin{array}{ll}
   \|\hat G^c(\x,\cdot) - \hat G^c_M(\x,\cdot)\|_{L^2}^2 
      &\leq  2\,e^{-2\,\alpha(M)\,|\x|}\,
           \int_M^\infty \frac{e^{-2\,A_0\,(\omega-M)\,|\x|}}{(4\,\pi\,|\x|)^2} \d \omega  \\
      &\leq \frac{e^{-2\,\alpha(M)\,|\x|}}{(4\,\pi\,|\x|)^2\,A_0\,|\x|}
\end{array}
$$
and 
$$ 
\begin{array}{ll}
   \|\hat G^c(\x,\cdot) \|_{L^2}^2 
      &\geq  \int_{-\infty}^\infty \frac{e^{-2\,(A_2\,|\omega| + A_1\,\omega^2)\,|\x|}}{(4\,\pi\,|\x|)^2} \d \omega \\
      &=  \frac{\sqrt{\pi}\,e^{\frac{A_2^2\,|\x|}{2\,A_1}}}{(4\,\pi\,|\x|)^2\,\sqrt{2\,A_1\,|\x|}}
\end{array}
$$
The theorem follows from~(\ref{resofPPE}) and the last two results. 
\end{pf}

\subsubsection{Remark.} Because $\omega\mapsto \hat G^c_M(\x,\omega)$ 
vanishes on a non-empty interval, causality condition~(\ref{causrequ}) 
is not satisfied for $G^c_M$. However, Theorem~\ref{theo:theo1} shows that  
$G^c_M(\x,t)$ is a good $L^2-$approximation of $G^c(\x,t)$ if $M$ is 
sufficiently large. \\

\begin{thm}\label{theo:theo2}  
Let $B_1:=\Re(\alpha_*^{pl} - \alpha_*^c)$, $B_2:=\Im(\alpha_*^{pl} - \alpha_*^c)$, 
$$
  C := \left|1 
               - 2\,e^{-B_1(\omega)\,|\x|}\,\cos(B_2(\omega)\,|\x|) 
               + e^{-2\,B_1(\omega)\,|\x|} \right|\,.
$$ 
and $M_\delta\in (0,M]$ be such that 
\begin{equation}\label{delta}
      \|\hat G^c_{M_\delta}(\x,\omega)\|_{L^2}^2 = (1-\delta) \|\hat G^c(\x,\omega)\|_{L^2}^2 \,
\end{equation}
for some $0<\delta<< 1$. For $|\x|\not=0$ it follows that
$$ 
\begin{array}{l}      
   \frac{\|G^c_M(\x,\cdot) - G^{pl}_M(\x,\cdot)\|_{L^2}}{\|G^c(\x,\cdot) \|_{L^2}} 
   \leq \sqrt{(1-\delta)\,D_1(|\x|) + \delta\,D_2(|\x|)} 
\end{array}
$$
with $D_j(|\x|):=\max_{\omega\in I_j} C(|\x|,\omega)^2$ for $I_1:=[-M_\delta,M_\delta]$ and 
$I_2:=\R\backslash I_1$. 
\end{thm}

\begin{pf}
As in the proof of Theorem~\ref{theo:theo1}, it follows that 
$$
  \frac{\|G^c_M(\x,\cdot) - G^{pl}_M(\x,\cdot)\|_{L^2}}{\|G^c_M(\x,\cdot) \|_{L^2}} 
   = \frac{\|\hat G^c_M(\x,\cdot) - \hat G^{pl}_M(\x,\cdot)\|_{L^2}}{\|\hat G^c_M(\x,\cdot) \|_{L^2}} \,. 
$$
The theorem follows from the last result,~(\ref{delta}) and 
$$
\begin{array}{lll}    
   &\int|\hat G^c_M(\x,\cdot) - \hat G^{pl}_M(\x,\cdot)|^2 \d\omega \\
     &\qquad\qquad\leq D_1(|\x|)\int_{|\omega|\leq M_\delta} |\hat G^c_M(\x,\cdot)|^2\d\omega \\
       &\qquad\qquad\quad +  D_2(|\x|) \int_{|\omega|\geq M_\delta} |\hat G^c_M(\x,\cdot)|^2\d\omega\,.
\end{array}
$$
Cf. Fig.~\ref{fig:fig03} from the example in Section~\ref{sec:NumEx} for which 
$\delta\approx 6\cdot 10^{-4}$, $M_\delta\approx 10\,MHz$, $D_1(|\x|)\approx 5.6\cdot 10^{-13}$ 
and $D_2(|\x|)\approx 1.04$, i.e. $\sqrt{(1-\delta)\,D_1(|\x|) + \delta\,D_2(|\x|)}=0.025$.
\end{pf}

\section{Numerical example}\label{sec:NumEx}

We now present a numerical example. We choose the following parameter values 
$$
   \gamma\approx 1.66,\quad c_0\approx 0.15\,\frac{cm}{\mu s},\quad 
   \tau_0 \approx 10^{-6}\,\mu s\,,
$$
$$
    \alpha_1 
      = \frac{2\,c_0\,a_1}{\tau_0^{\gamma-1}\,|\cos(\gamma\,\pi/2)|} 
      \approx 138.08\,MHz
$$
(where we have used~(\ref{a1a2}) and $a_1\approx 0.04344\,\frac{1}{cm}$) and 
$$
    a_2 = \frac{\alpha_1}{c_0}\approx 920.55\,\frac{1}{cm} \,
$$
to model a medium similar to \emph{castor oil}. (For the relaxation time $\tau_0$ 
see~\cite{KiFrCoSa:00} and for $a_1$ and $\gamma$ see~\cite{Sz:95}). 

\begin{figure}[!ht]
\begin{center}
\includegraphics[height=5.0cm,angle=0]{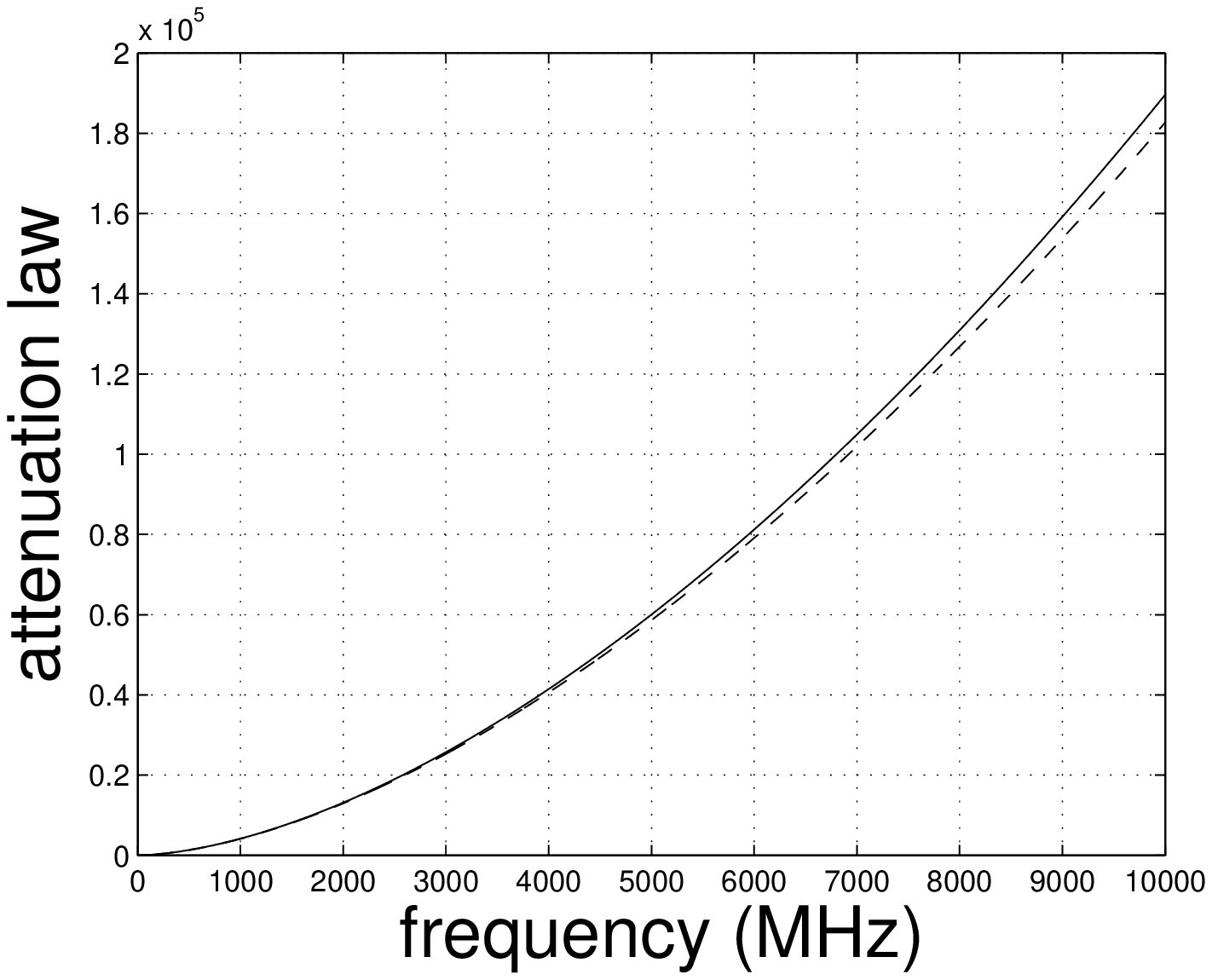}
\includegraphics[height=5.0cm,angle=0]{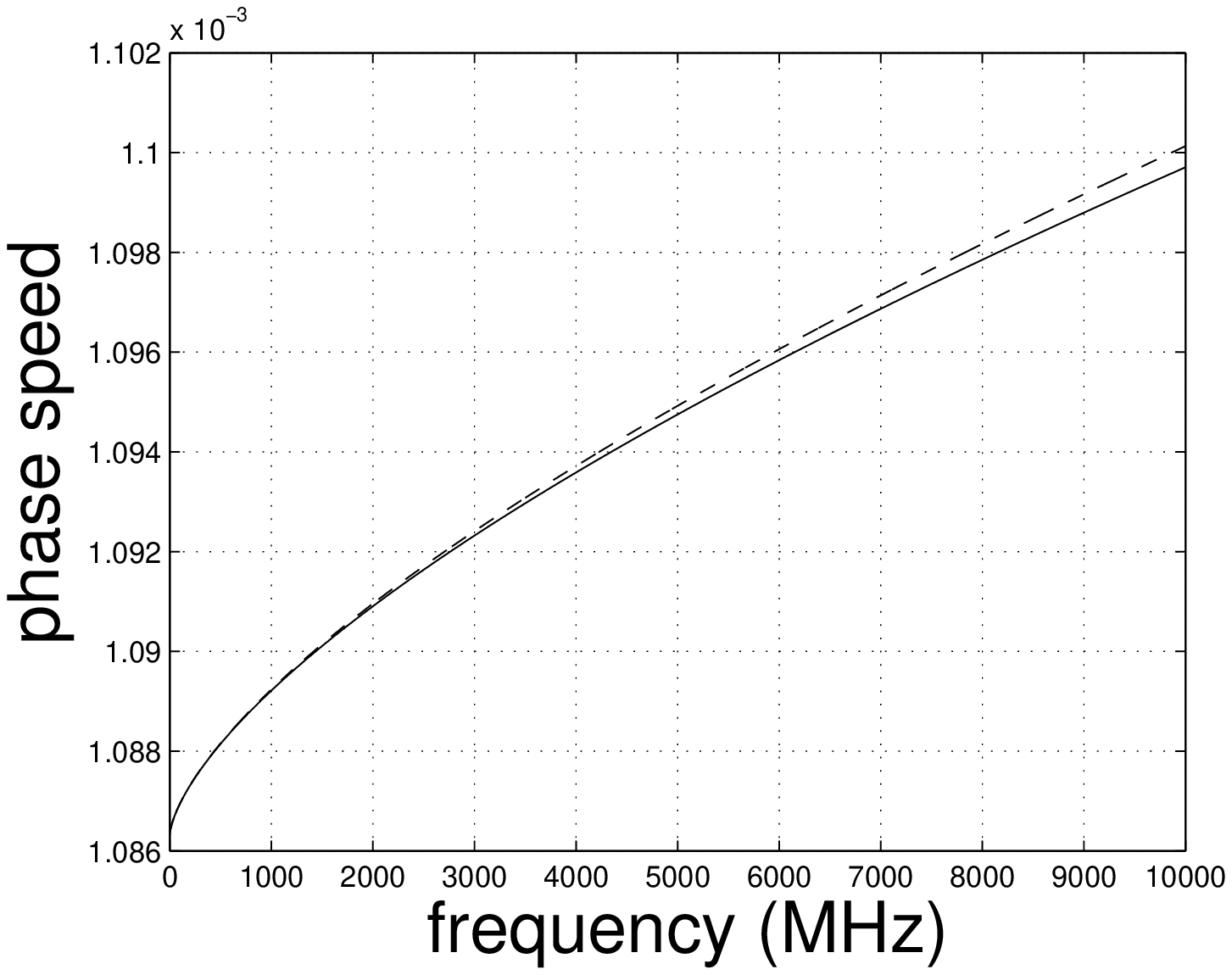}
\end{center}
\caption{Numerical comparison of the attenuation laws and phase speeds. 
The dashed lines correspond to model $\alpha_*^c(\omega)$ and the solid 
lines correspond to model $\alpha_*^{pl}(\omega)$. 
}\label{fig:fig01}
\end{figure}

The differences of the attenuation laws and the phase speeds of the attenuation-dispersion 
laws $\alpha_*^c$ and $\alpha_*^{pl}$ are visualized in Fig.~\ref{fig:fig01}. 
From the figure we see that $\alpha_*^{pl}$ is a good approximation of the attenuation-dispersion 
law $\alpha_*^c$ for the (practically measurable) frequency range $[0,60]\,MHz$. In the following 
we choose as small frequency bound: 
$$
           M=100\,MHz << \frac{1}{\tau_0}\,.
$$

\subsubsection{Small frequency approximation of causal model.} For the estimation in 
Theorem~\ref{theo:theo1} we choose the following constants
$$
   A_0= 0.7\,\frac{\d \alpha^{pl}}{\d \omega}(M),\quad
   A_1=a_1 \quad\mbox{and}\quad A_2=a_2\,
$$
and obtain 
$$ 
   \frac{\|G^c(\x,\cdot) - G^c_M(\x,\cdot)\|_{L^2}}{\|G^c(\x,\cdot) \|_{L^2}} 
   \leq 0.0828\,
            \frac{e^{-4.877\cdot 10^6\cdot |\x|}}{\sqrt[4]{|\x|}} \,.
$$
Therefore the frequency range $[-10^2,10^2]$ is reasonable for a small frequency approximation 
of $G^c$ for distances larger than $10^{-6}\,cm$.

\subsubsection{Small frequency approximation via power law.} 
In Fig.~\ref{fig:fig03} we have visualized the functions $M_0\mapsto g(M_0):= \|G_{M_0}^c(\x,\cdot)\|_{L^2}$ 
and $\omega\mapsto C(|\x|,\omega)$ for $|\x|=1\,cm$. According to this figure and Theorem~\ref{theo:theo2}, 
$G_M^{pl}$ is a good $L^2-$approximation of $G^c$ for the case $|\x|=1\,cm$. 
In Table~\ref{tb:epsM} we have listed the relative error
$$ 
  \epsilon_M(|\x|) 
    := \frac{\|G_M^c(\x,\cdot) - G_M^{pl}(\x,\cdot)\|_{L^2}}{\|G_M^c(\x,\cdot)\|_{L^2}}\,,
$$
for different distances $L$.  
\begin{table}[ht]
\begin{center}
\caption{Visualization of error $\epsilon_M$ for different distances $L$ (in $cm$).}
\label{tb:epsM}
\begin{tabular}{ccccc}
\hline
$L$          & $10^{-6}$ & $10^{-3}$ & $10^{-1}$ & $10$                    \\
$\epsilon_M$ & $7.62\cdot 10^{-8}$  & $7.35\cdot 10^{-5}$ & $4.46\cdot 10^{-4}$  & $7.13\cdot 10^{-5}$  \\ \hline
\end{tabular}
\end{center}
\end{table}
We see that $G_M^c(\x,t)$ and $G_M^{pl}(\x,t)$ are good approximations of 
$G^c(\x,\cdot)$ as long as 
$$
     M\geq 100\,MHz  \quad\mbox{and}\quad 
     |\x| \geq 10^{-6}\,cm\,.
$$
In particular, this means that the wave can be predicted for  
$$
      t \geq \frac{10^{-6}\,cm}{c_0} = 6.67\cdot 10^{-6}\,\mu\,s\,.
$$

\subsubsection{Remark on causality.} 
Because the wave $G^c$ has the finite wave front speed $c_F=c_0$ and $G^{pl}_M(\x,\cdot)$ is 
a good $L^2-$approximation of $G^c(\x,\cdot)$, it follows that $G^{pl}_M$ has in ``some sense'' 
the same wave front speed. 
In ``some sense'' means that $G^{pl}_M$ contains partial waves propagating arbitrarily fast but 
their sum is small in the $L^2-$sense.

\subsubsection{Remark on large frequencies.} 
\begin{figure}[!ht]
\begin{center}
\includegraphics[height=5.0cm,angle=0]{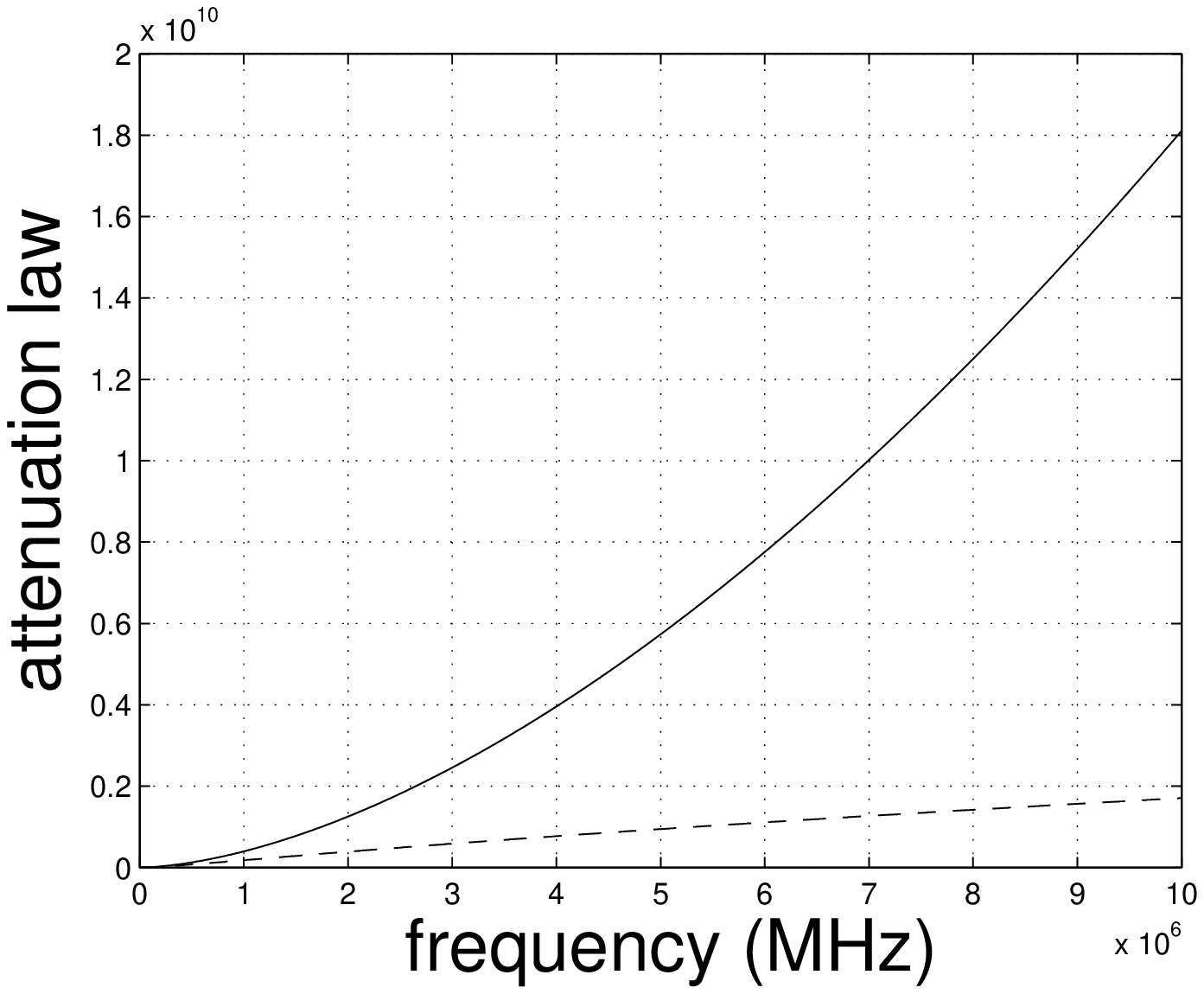}
\includegraphics[height=5.0cm,angle=0]{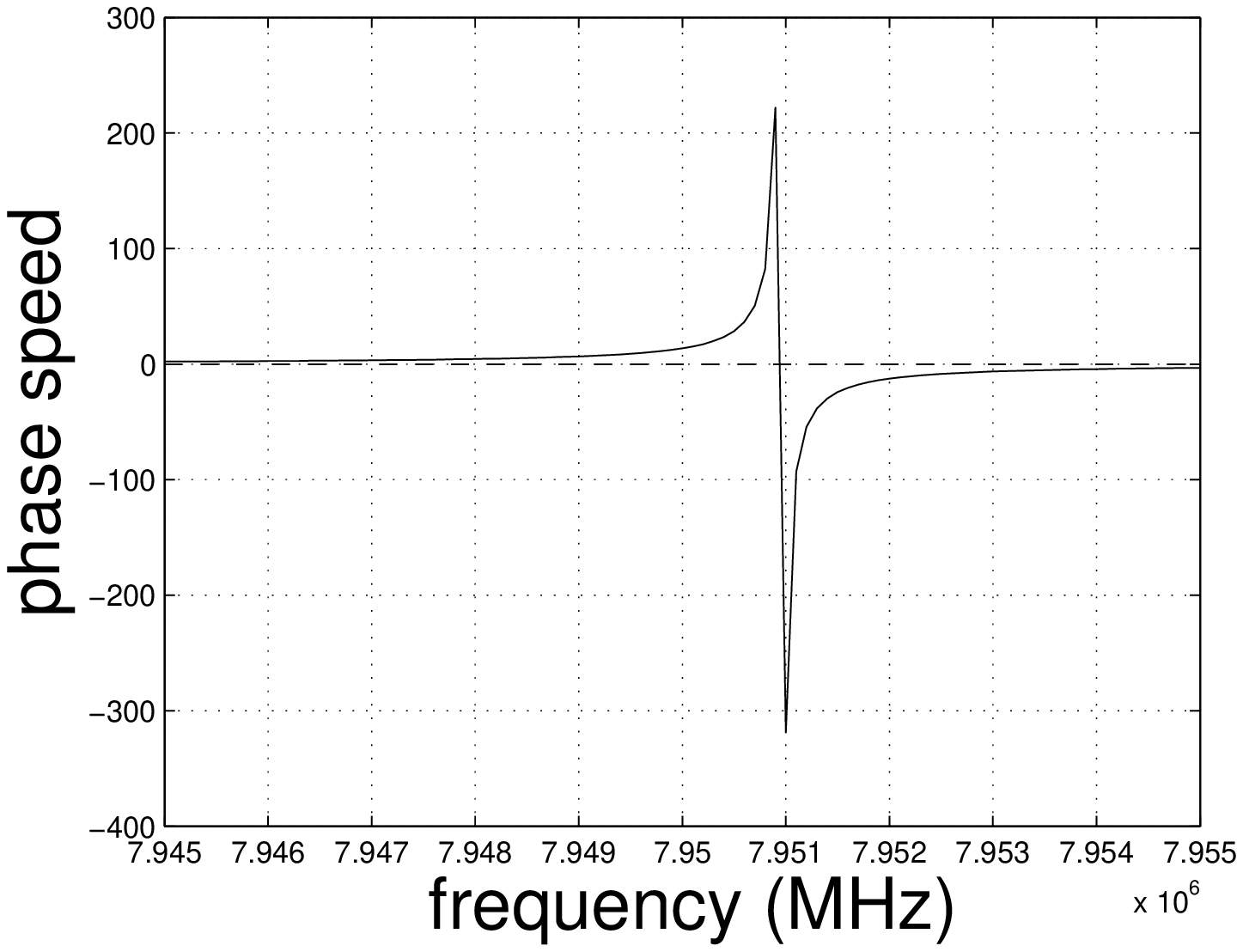}
\end{center}
\caption{Numerical comparison of the attenuation laws and phase speeds for large frequencies  
The dashed lines correspond to model $\alpha_*^c(\omega)$ and the solid 
lines correspond to model $\alpha_*^{pl}(\omega)$. 
}\label{fig:fig02}
\end{figure}

\begin{figure}[!ht]
\begin{center}
\includegraphics[height=5.0cm,angle=0]{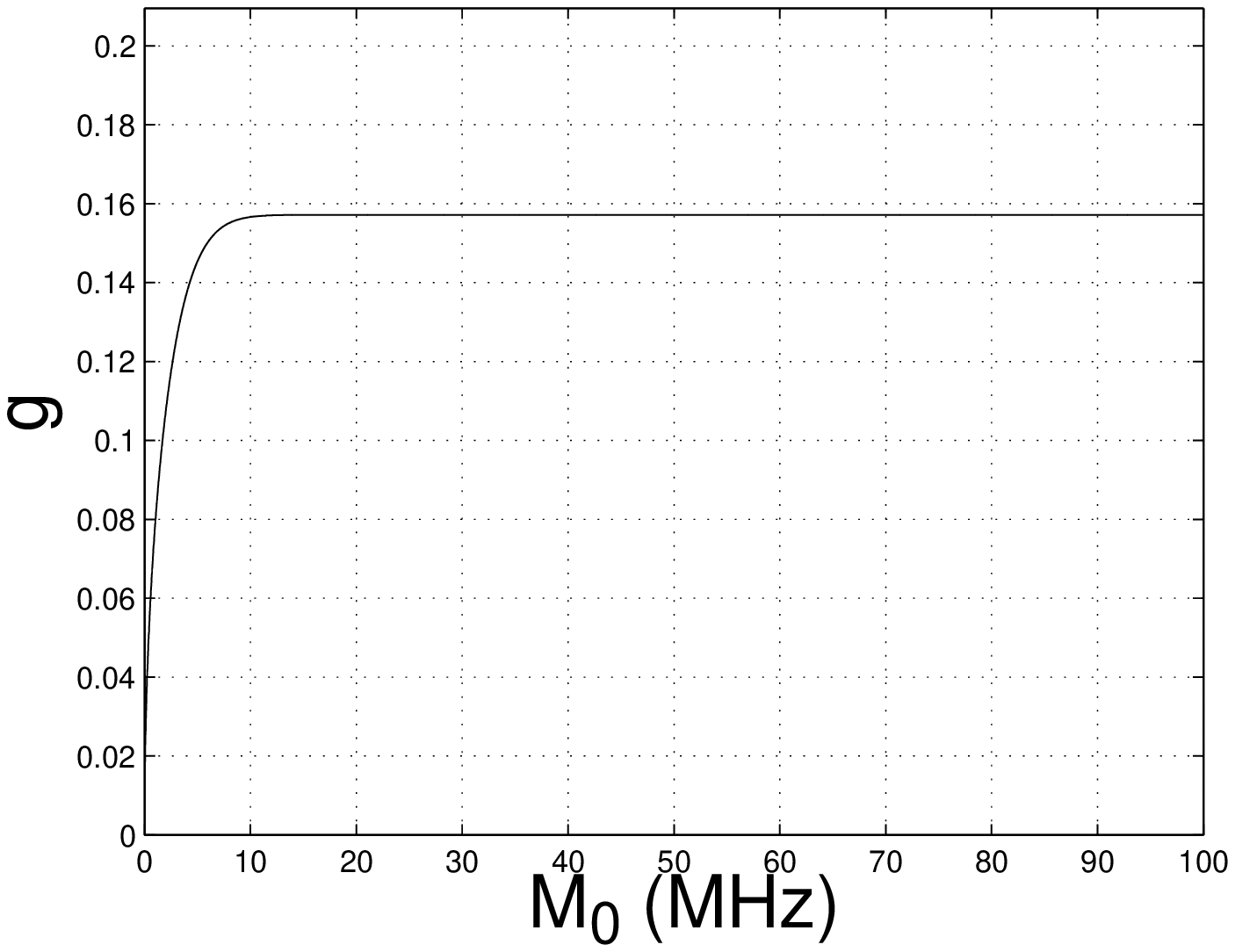}
\includegraphics[height=5.0cm,angle=0]{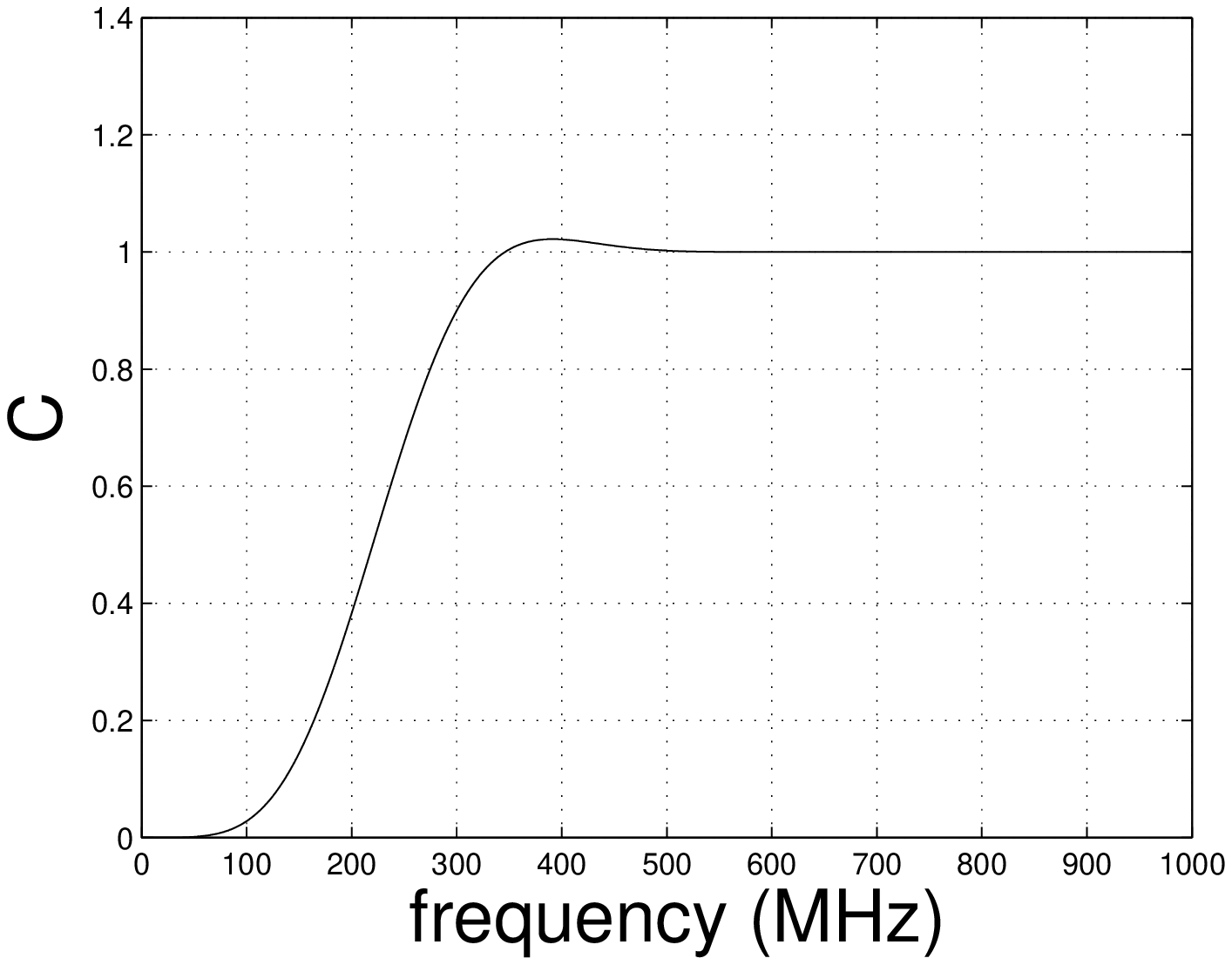}
\end{center}
\caption{Visualization of functions $M_0\mapsto g(M_0):= \|G_{M_0}^c(\x,\cdot)\|_{L^2}$  
and $\omega\mapsto C(|\x|,\omega)$ for $|\x|=1\,cm$. 
}\label{fig:fig03}
\end{figure}

We note that the power attenuation law 
$\alpha^{pl}(\omega)$ increases much faster than $\alpha^c(\omega)$ for large frequencies 
and that the respective phase speed $c^{pl}(\omega)$ has a singularity  
at $\omega_1\approx 7.950959\cdot 10^6\,MHz$ and at $\omega_2=-\omega_1$ (cf.~Fig~\ref{fig:fig02}), 
respectively. 
In contrast to the latter fact, the phase speed $c^c(\omega)$ of the causal model has no singularity.

\section{Conclusions}

We showed that - for appropriate situations - the causal 
wave $G^c$ defined by~(\ref{green}),~(\ref{beta}) and~(\ref{alpha*}) 
can be approximated by the noncausal wave $G^{pl}_M$ defined by~(\ref{defhatGM})~(\ref{green}),
~(\ref{beta}),~(\ref{alpha*pl}) and~(\ref{a1a2}). In words, the wave $G^{pl}$ contains 
partial waves propagating arbitrarily fast but the sum of the noncausal waves is small 
in the $L^2-$sense. 
This result fits in with the results in~\cite{KeMcMe:08}.

\end{document}